\documentclass[aps,pra,showpacs,amsmath,amssymb,twocolumn]{revtex4-1}
\usepackage{graphicx}
\usepackage[english]{babel}

\def \bfm #1{\mbox{\boldmath{$ #1 $}}}

\newcommand{\mumeter}[0]{\ensuremath{\mu\mathrm{m}}}

\newcommand{\Al}[0]{\ensuremath{{^{27}\mathrm{Al}^+}}}
\newcommand{\Ca}[0]{\ensuremath{{^{40}\mathrm{Ca}^+}}}
\newcommand{\Be}[0]{\ensuremath{{^{9}\mathrm{Be}^+}}}
\newcommand{\Nitrogen}[0]{\ensuremath{{^{14}\mathrm{N}_2^+}}}
\newcommand{\Oxigen}[0]{\ensuremath{{^{16}\mathrm{O}_2^+}}}
\newcommand{\ket}[1]{\ensuremath{|{#1}\rangle}}
\newcommand{\bra}[1]{\ensuremath{\langle{#1}|}}
\newcommand{\threejm}[6]{ \left(\begin{array}{ccc} #1 & #3 & #5\\
      #2 & #4 & #6
    \end{array}
  \right)}
\newcommand{\sixj}[6]{ \left\{\begin{array}{ccc} #1 & #3 & #5\\
      #2 & #4 & #6
    \end{array}
  \right\}}
\newcommand{\ninej}[9]{\left\{\begin{array}{ccc}
      #1 & #2 & #3\\
      #4 & #5 & #6\\
      #7 & #8 & #9
    \end{array}
  \right\}}

\begin{document}

\title{Temperature-independent quantum logic for molecular spectroscopy}

\author{Jordi Mur-Petit}
\email{Corresponding author: jordi.mur@iff.csic.es}
\affiliation{Instituto de F\'\i sica Fundamental, IFF-CSIC, Serrano 113 bis,
  E-28006 Madrid, Spain}
\author{Jes\'us P\'erez-R\'\i os}
\affiliation{Instituto de F\'\i sica Fundamental, IFF-CSIC, Serrano 123,
  E-28006 Madrid, Spain}
\author{Jos\'e Campos-Mart\'\i nez}
\affiliation{Instituto de F\'\i sica Fundamental, IFF-CSIC, Serrano 123,
  E-28006 Madrid, Spain}
\author{Marta I. Hern\'andez}
\affiliation{Instituto de F\'\i sica Fundamental, IFF-CSIC, Serrano 123,
  E-28006 Madrid, Spain}
\author{Stefan Willitsch}
\affiliation{Department of Chemistry, University of Basel,
  Klingelbergstrasse 80, CH-4056 Basel, Switzerland}
\author{Juan Jos\'e Garc\'\i a-Ripoll}
\affiliation{Instituto de F\'\i sica Fundamental, IFF-CSIC, Serrano 113 bis,
  E-28006 Madrid, Spain}



\begin{abstract}
We propose a fast and non-destructive spectroscopic method for single molecular ions that implements quantum logic schemes between an atomic ion and the molecular ion of interest. Our proposal relies on a hybrid coherent manipulation of the two-ion system, using optical or magnetic forces depending on the types of molecular levels to be addressed (Zeeman, rotational, vibrational or electronic degrees of freedom). The method is especially suited for the non-destructive precision spectroscopy of single molecular ions, and sets a starting point for new hybrid quantum computation schemes that combine molecular and atomic ions, covering the measurement and entangling steps.
\end{abstract}

\maketitle

\section{Introduction}

Recent advances in trapping and cooling of molecules and molecular ions are opening new perspectives in fields as varied as cold chemistry, quantum dynamics of complex systems, metrology and quantum computation~\cite{carr2009,bell09b, flambaum07,demille02a}. In this context, the sympathetic cooling of molecular ions represents a particularly attractive approach which enables to perform chemical and spectroscopic experiments on the single-particle level~\cite{stefanpccp}. With the recent breakthrough in the production of cold molecular ions in well-defined internal states~\cite{MhH+-natp10,HD+-natp10,tong2010,tong2011}, fully coherent experiments with single molecular ions now come within reach. Precise knowledge and control of the internal quantum state of molecular ions is relevant for quantum controlled chemistry~\cite{stefanpccp,ospelkaus10b}, frequency metrology of single ions \cite{schmidt05, schiller05a} and prospective applications in quantum-information processing, where a lot of progress has already been made with trapped atomic ions~\cite{leibfried03-rmp,haeffner2008}. However, readout procedures of the internal quantum state have thus far relied on methods such as laser-induced charge transfer~\cite{ tong2010,tong2011} or photodissociation~\cite{MhH+-natp10,HD+-natp10} because of the lack of closed cycling transitions in molecular systems. These techniques are inherently destructive and not suitable if repeated measurements on a single molecular ion are required, for instance in quantum computation~\cite{haeffner2008} or frequency metrology~\cite{schmidt05}.

In this work we propose a fast, efficient and accurate molecular spectroscopy scheme based on quantum logic and  coherent control theory, able to address a variety of internal degrees of freedom of the molecules. Our proposal relies on a hybrid manipulation of the ion of interest and an atomic ion, such as \Ca\ or \Be, that acts as a probe. Optical or magnetic forces are used on the ions at will, depending on the precise internal states which are involved in the quantum operations. They are arranged in optimized gates that take a time from $50\,\mu\mathrm{s}$ up to $1$ ms, are insensitive to the temperature of the ion crystal and have a number of relevant applications. The most immediate one is the determination of magnetic moments or Zeeman shifts in atoms and molecules, such as \Nitrogen\ or \Oxigen, for precision-spectroscopic purposes. This pushes the field along the line of previous experiments with atomic ions~\cite{schmidt05}, incorporating two important features: a greater versatility (that allows addressing a broader variety of ions with a complex internal structure), and simpler experimental requirements, as cooling the ions to the motional ground state is not necessary. Another application of this work is to replace the destructive measurement techniques in current experiments with hybrid atom-molecule systems~\cite{tong2010,tong2011}. Moreover, the proposed framework represents the corner stone of a hybrid quantum computation scheme which for the first time combines molecular and atomic ions, covering the measurement and entangling steps. Finally, in the broader context of molecular physics, the most important aspect of the scheme is that it represents {\em an entirely new approach for molecular spectroscopy} which relies on the manipulation of quantum phases and is tailored towards the interrogation of single trapped particles.

Our manuscript is structured as follows. We start in Sect.~\ref{sec:qls} by reformulating the concept of quantum logic spectroscopy~\cite{schmidt05}, showing how by means of control-phase gates and Ramsey interferometry, a single ion may probe the state population or other observables of any other ion in a Coulomb crystal~\cite{stefanpccp}. We then show how to implement those control-phase gates using state-dependent forces~\cite{garcia-ripoll05,garcia-ripoll03,leibfried03} in a crystal with different types of atomic and molecular ions, thus introducing a quantum protocol for molecular ions. With these tools, in Sect.~\ref{sec:implementation} we derive three slightly different variants of this protocol, and test them with accurate calculations for the case of \Nitrogen\ ions. In the first version, the probe and the target ions are subject to pulsed optical forces generated by AC Stark shifts. This fast protocol is targeted at the determination of electronic, rotational and vibrational states of the target molecular ion. In the second version, the ions are subject to oscillating magnetic fields, allowing the determination of Zeeman states. The final version relies on femtosecond laser pulses acting on the control ion, leading to an order of magnitude greater sensitivity of the detection scheme. We draw our conclusions in Sect.~\ref{sec:conclusions}.

\section{Quantum logic spectroscopy}
\label{sec:qls}

Roughly speaking, spectroscopy studies the reaction of a physical system under periodic drivings, forces, or radiation of various frequencies. The most common approach is to watch the back-action of the system on the radiation that drives it, measuring the absorption, the emission or the phase in those fields. The field of quantum logic spectroscopy (QLS), introduced in Ref.~\cite{schmidt05}, advocates a very accurate study of the changes suffered by the driven system, using quantum gates to enhance the precision of those studies. In this section we extend these ideas to the field of molecular spectroscopy. To this end, we have to overcome the two main limitations of the original QLS protocol, which are the temperature requirements and its speed.

\subsection{Phase sensitive QLS protocol}
\label{ssec:phase-qls}

The original experiment for QLS~\cite{schmidt05} employed two atomic ions that allow optical manipulation, \Be\ and \Al. By means of sideband transitions, the state of the target ion, \Al, which cannot be directly measured with spectroscopic accuracy because it lacks cycling transitions, was mapped onto the control ion, \Be, which was later on measured using the accurate technique of electron shelving. Hence, in this pioneering work, quantum logic addressed the lack of cycling transitions in \Al, at the price of imposing accurate coherent operations (sidebands) and ground state cooling of the motional state of the ions in the trap. Both requirements are experimentally demanding and have been addressed in later experiments with atomic ions (see e.g.~\cite{hermann2009,clark2010}). However, they have not yet been achieved for sympathetically-cooled molecular species.

\begin{figure}[t]
  \centering
  \includegraphics[width=\linewidth]{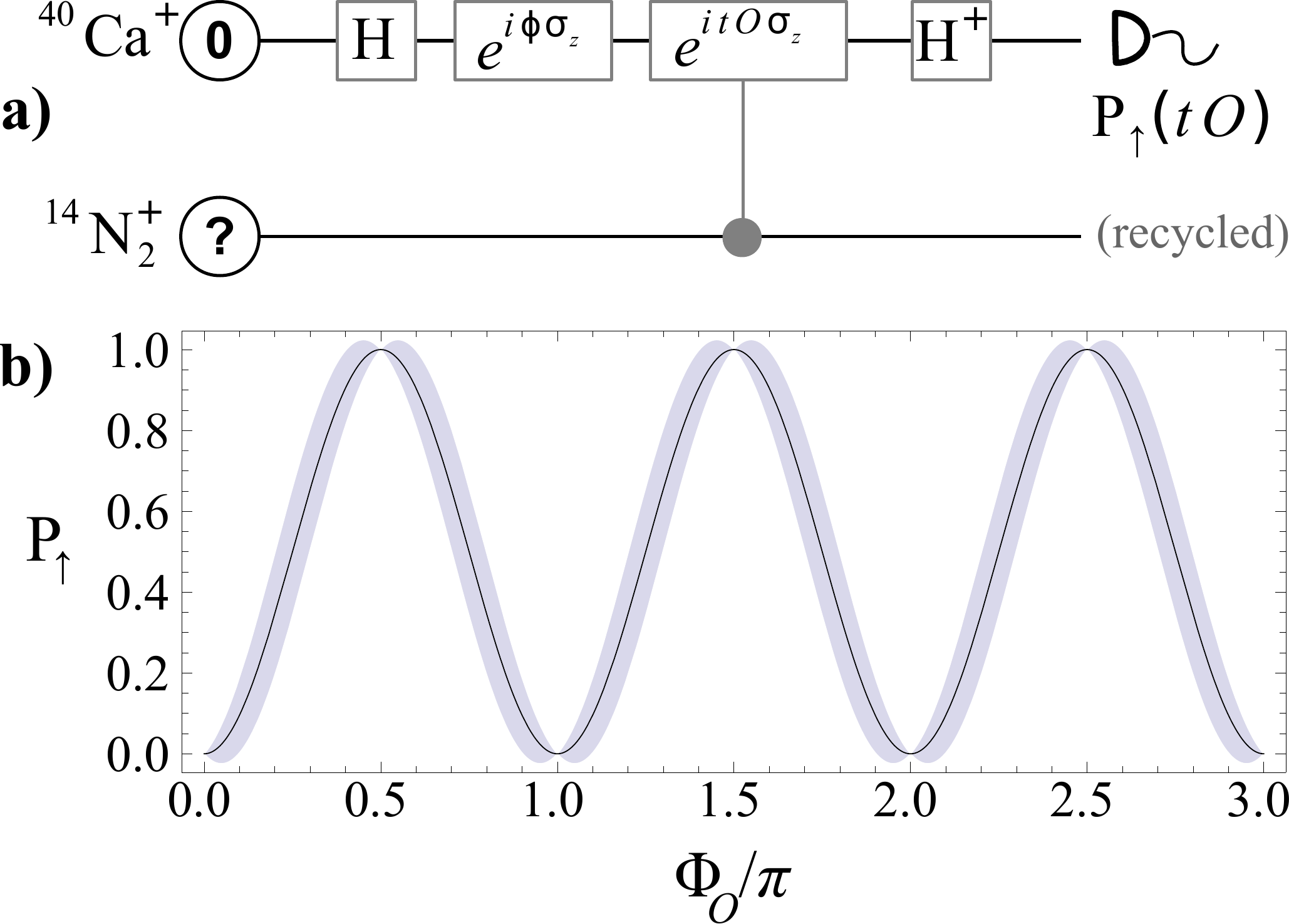}  
  \caption{State detection protocol. (a) The probe ion experiences a series of local gates combined with a two-qubit control-phase gate in which the control is a property, $O,$ of the target. (b) Oscillations in the state of the probe ion as a function of the acquired phase, $\Phi_O \propto t O.$ We plot the mean excitation probability with the error after 10 repetitions $(\xi=0).$}
  \label{fig:protocol}
\end{figure}

One can dramatically enhance the versatility of the QLS protocol by using geometric gates based on state-dependent forces~\cite{leibfried03,garcia-ripoll03,garcia-ripoll05}. These are quantum gates that are insensitive to the motional state of the ions and may thus be applied to sympathetically cooled molecular ions. The underlying physical principle of these gates, explained in Sect.~\ref{ssec:geometric-phase}, is that when a chain of ions is shaken by some external forces, the quantum state acquires a phase that depends on those forces. Moreover, when the forces experienced by the ions are ``state-dependent'', that is, when they have different magnitudes or signs depending on the internal state of the ions, then the quantum phase carries precise information on the state of those atomic or molecular ions. This information is the one we use for the QLS algorithm.

More precisely, assume that we have two ions, a control or logic ion, labeled C, and a target or spectroscopy ion, labeled T, confined in the same ion trap. The C ion is Doppler laser cooled and sympathetically cools the target ion to form a two-ion Coulomb crystal. Both ions will be subject to independent external forces, different from the laser-cooling process.  The ion C will be a qubit, and the force acting on it will depend on its internal state, $f_{\textrm{C}}(t)\sigma^z_{\textrm{C}},$ while the force on T will depend on some property of this ion, $f_{\textrm{T}} (t) O,$ such as a magnetic moment or a quantum number. Here, $\sigma^z_{\textrm{C}}$ is the Pauli $z$-matrix for the internal state of the C ion. As explained in Sect.~\ref{ssec:geometric-phase}, the quantum state of the system will acquire a total nontrivial phase of the form
\begin{equation}
  \label{eq:phase-gate}
  \Phi_O = \sigma^z_{\textrm{C}} O \phi_{\textrm{CT}},
\end{equation}
where $\phi_{\textrm{CT}}$ is a function of the normal modes of the two ions, $\omega_{\textrm{com}}$ and $\omega_{\textrm{str}}$~\footnote{Note the different ions experience different trapping frequencies, and this influences their eigenmodes, $\omega_{\textrm{com,str}}$}, and their combined drivings. This phase is robust, it does not depend on the motional state of the ions at the beginning of the gate operation and thus is independent of temperature. Based on this we design a quantum protocol which interrogates very accurately the state-dependent phase $\Phi_O$, obtaining the value of the observable $O$ (cf.~Fig.~\ref{fig:protocol}a)
\begin{enumerate}
\item Prepare ion C in state $\ket{0};$
\item apply a Hadamard gate on C, $H=\exp(-i\sigma^y_{\textrm{C}}\pi/2),$
\item optionally, apply a reference phase on the control ion, $\exp(i\phi\sigma^z_{\textrm{C}});$
\item apply the state dependent forces $f_{\textrm{C,T}} (t)$ on the C and T ions, ensuring that the initial motional state of the ions is restored (see Sect.~\ref{ssec:zeeman});
\item undo the Hadamard gate, $H^\dagger;$ and
\item measure the state of the control ion.
\end{enumerate}
At the end of this process the probe and the target ions will get entangled and the excited state population of the control ion will oscillate as (cf.~Fig.~\ref{fig:protocol}b)
\begin{equation}
  P_\uparrow = \sin^2( t O \phi_{\textrm{CT}} + \xi).
\end{equation}
This allows, by repeated application of the protocol, to determine the value of $O$ with very high precision~\cite{hume2007}.

We have so far presented the protocol in a very general form, without clarifying the origin and the time dependence of the forces that act on the control and target ion. In the following section we will elaborate on the physics behind the phase gate~(\ref{eq:phase-gate}), developing the formalism to compute and optimize the control and target forces. Using this we will then work out the actual implementation of the protocol to detect molecular electronic, rovibrational or Zeeman states, using optical or magnetic fields.

\subsection{Geometric phases}
\label{ssec:geometric-phase}

As pointed out in Ref.~\cite{leibfried03}, quantum harmonic oscillators possess a very simple geometric structure: a continuous displacement of the oscillator through phase space induces a global phase on the wavefunction, and this phase only depends on the initial and final points of the path, i.e., it is geometric in nature. Following the framework in Ref.~\cite{garcia-ripoll05}, one can show that when a harmonic oscillator experiences a force $\hbar f(t)$ for a time $T$, its quantum state acquires a phase
\begin{equation}
  \phi[\omega,f] = a_0^2\int_0^T\int_0^{\tau_1}\sin[\omega(\tau_1-\tau_2)] f(\tau_1)f(\tau_2) d\tau_2 d\tau_1,
\end{equation}
where $\omega$ and $a_0$ are the harmonic oscillator frequency and length, respectively. This phase is proportional to the area in phase space covered by a coherent state subject to this force, whose trajectory is
\begin{equation}
  z(t;\omega,f) = z(0) - i \frac{a_0}{\sqrt{2}} \int_0^t f(\tau)e^{i\omega \tau} d\tau.
\end{equation}
This geometric phase is resilient to temperature, for the area does not depend on the initial state of the oscillator, $z(0).$ Moreover, it is also insensitive to random errors in $f(t),$ which only appear as second-order corrections.

In our protocol we will act with two independent forces, $f_{\mathrm{C}}$ and $f_{\mathrm{T}}$, on two ions. In this case we have not one but two harmonic oscillators, associated to the center of mass (com) and stretch mode (str) of the ion crystal. Since the target and control ions are different, they will in general have different masses and also experience different confinement potentials. This complicates the final expressions, but we can still write the effective Hamiltonian
\begin{equation}
  H = \sum_{s\in\{\mathrm{com},\mathrm{str}\}} 
  \left\{
    \hbar\omega_{s} a^\dagger_s a_a + \hbar F_s(t) \frac{1}{\sqrt{2}}(a_s + a^\dagger_s)
  \right\},
\end{equation}
where $F_{\mathrm{com}}$ and $F_{\mathrm{str}}$ are linear combinations of $\sigma^z_{\mathrm{C}} f_{\mathrm{C}}(t)$ and $O f_{\mathrm{T}}(t)$ (cf. Appendix~\ref{sec:app-phase}). The total phase acquired by the combined system may be written
\begin{equation}
  \Phi = \phi[\omega_{\mathrm{com}},F_{\mathrm{com}}] + \phi[\omega_{\mathrm{str}},F_{\mathrm{str}}].
  \label{eq:totalphase}
\end{equation}
In this expression there will be many contributions, but the only one that will influence the signal of our protocol is the one proportional to $\sigma^z_{\mathrm{C}} \times O$, for it is the only entangling phase in this process. This nonlocal phase may be rewritten as the product (\ref{eq:phase-gate}), with some integral $\phi_{\mathrm{CT}}$ that depends on the actual driving, but which may be computed analytically or numerically.

\section{Implementation of the QLS protocol}
\label{sec:implementation}

In the following sections we will present three different versions of the protocol that can use different types of forces and drivings, and how they can be used to probe different properties of the target ion. In order to derive analytical estimates, we will work with Gaussian pulses having a duration $T$ and driven at frequency $\nu$ 
\begin{equation}
  f_{\textrm{C,T}} (t) = f_{\textrm{C,T}}^0\, e^{-(2t/T)^2} \cos(\nu t).\label{eq:gaussian}
\end{equation}
This ansatz does not represent any loss of generality, as the quantum gates may be further optimized choosing more sophisticated dependencies. Nevertheless, experience shows that the limits set by the previous pulses are very close to those achievable by more general forces~\cite{garcia-ripoll03}.

For this time dependence of the driving, and a sufficiently long pulse, $T\ge 5\pi/\mathrm{min\{|\nu-\omega_{com}|,|\nu-\omega_{str}|\}}$ so that {\em all} motional degrees of freedom are restored to their initial state, the two-ion state accumulates a total phase as given by Eq.~\eqref{eq:phase-gate} with (cf. Appendix~\ref{sec:app-phase})
\begin{align}
  \phi_{\textrm{CT}} &= \frac{1}{4}\sqrt{\frac{\pi}{2}} \frac{ f_{\textrm{C}}^0 f_{\textrm{T}}^0 a^2 T}{\omega} \Xi(\nu,\omega,\mu) .
  \label{eq:phaseCT}
\end{align}
Here we introduced the enhancement function
\begin{align}
  \Xi (\nu,\omega,\mu) &:=
      \frac{\omega^2}{\omega^2_{\textrm{com}}(\mu)-\nu^2} - \frac{\omega^2}{\omega^2_{\textrm{str}}(\mu)-\nu^2} ,
\end{align}
where
$\mu=m_{\textrm{T}}/m_{\textrm{C}}$ is the mass ratio of the two ions, 
$a^2 = \hbar/[(m_{\textrm{C}}+m_{\textrm{T}})\omega]$, 
and $\omega$ is the frequency of the control ion trap.
Note that our formalism allows the study of arbitrary drivings, $\nu$, which can be advantageous in some setups, as we discuss below.

\subsection{Optical kicking protocol to address electronic and rovibrational states}
\label{ssec:elec}

\begin{figure}
  \centering
  \includegraphics[width=\linewidth]{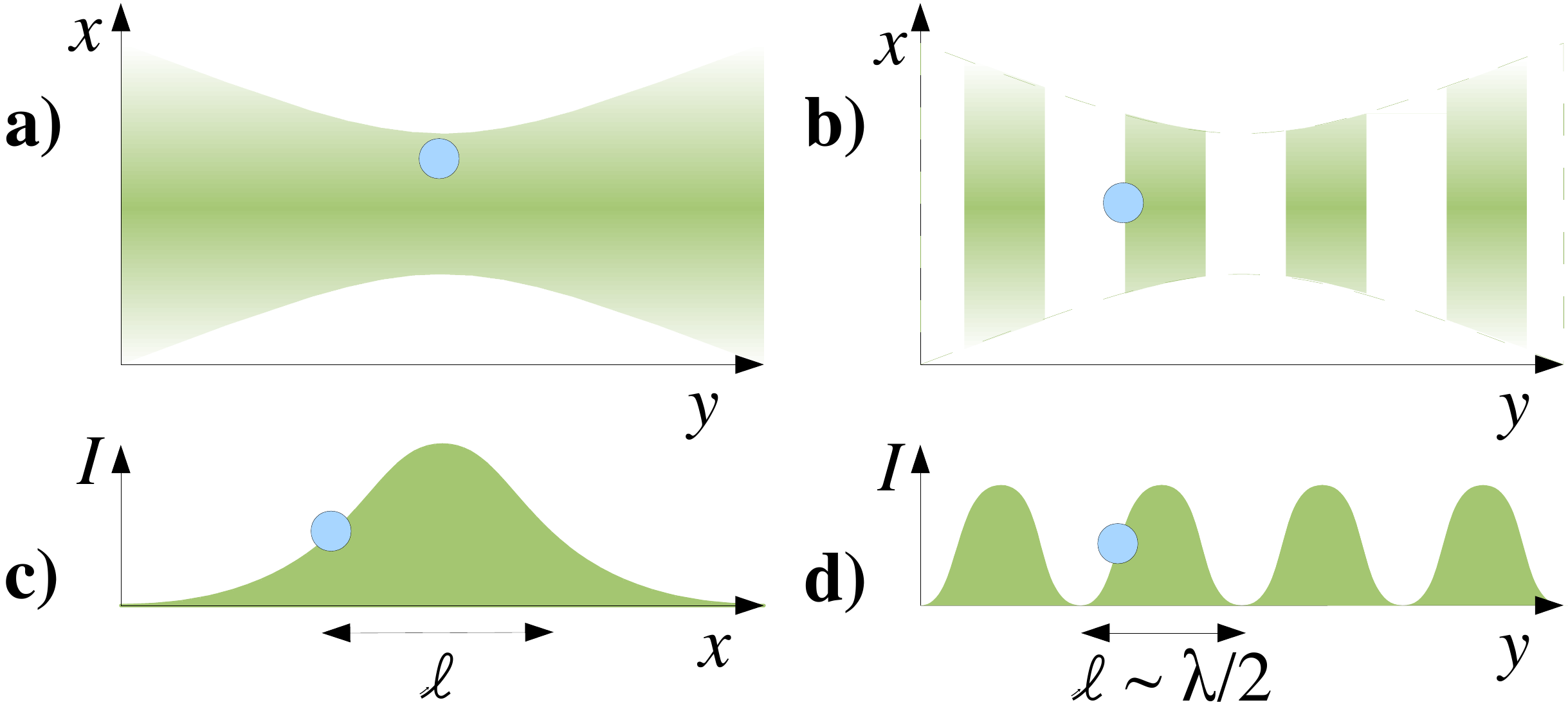}
  \caption{(color online) Scheme for imparting an optical force on an ion. (a,c) A single off-resonant laser beam is focused on the particle with a radial displacement, so as to achieve the maximum intensity gradient. (b,d) Two counter-propagating laser beams create an optical lattice and the atom sits on the point of maximum slope.}
  \label{fig:scheme}
\end{figure}

The first version of the protocol aims at distinguishing electronic, vibrational and rotational states of the molecular ion. For these degrees of freedom it is both advantageous and very efficient to rely on optical ``kicks'' as the state-dependent driving mechanism behind the gate. However, due to the inherent complexity of the molecular energy-level structure, we must seek gentle ways to drive the charged particles without accidentally exciting any undesired internal transitions. For this purpose, we will employ the dipole forces induced by two lasers that are detuned from an electronic transition of the atomic and the molecular ion, respectively. Each ion will sit on a spot of maximum slope of intensity, experiencing an AC Stark shift force
\begin{equation}
  {f}_{\textrm{C,T}}(t) = \frac{\Omega(t)^2}{\Delta} \times \frac{1}{\ell},
\end{equation}
where $\Delta$ is the laser detuning, $\Omega$ is the Rabi frequency and $\Omega^2$ is proportional to the light intensity~\footnote{For simplicity, we assume two independent lasers, each one acting either on the C or the T ion, with the same Rabi frequency and detuning, but not phase locked. However, none of these assumptions is essential for the results.}. The force depends on some characteristic length over which the light intensity varies, $\ell$, which ranges from half a wavelength for an optical lattice, up to $\ell \sim 5 - 100 \mumeter$ for a focused laser beam~(see Fig.~\ref{fig:scheme}).

\begin{figure}
  \centering
  \includegraphics[width=0.8\linewidth]{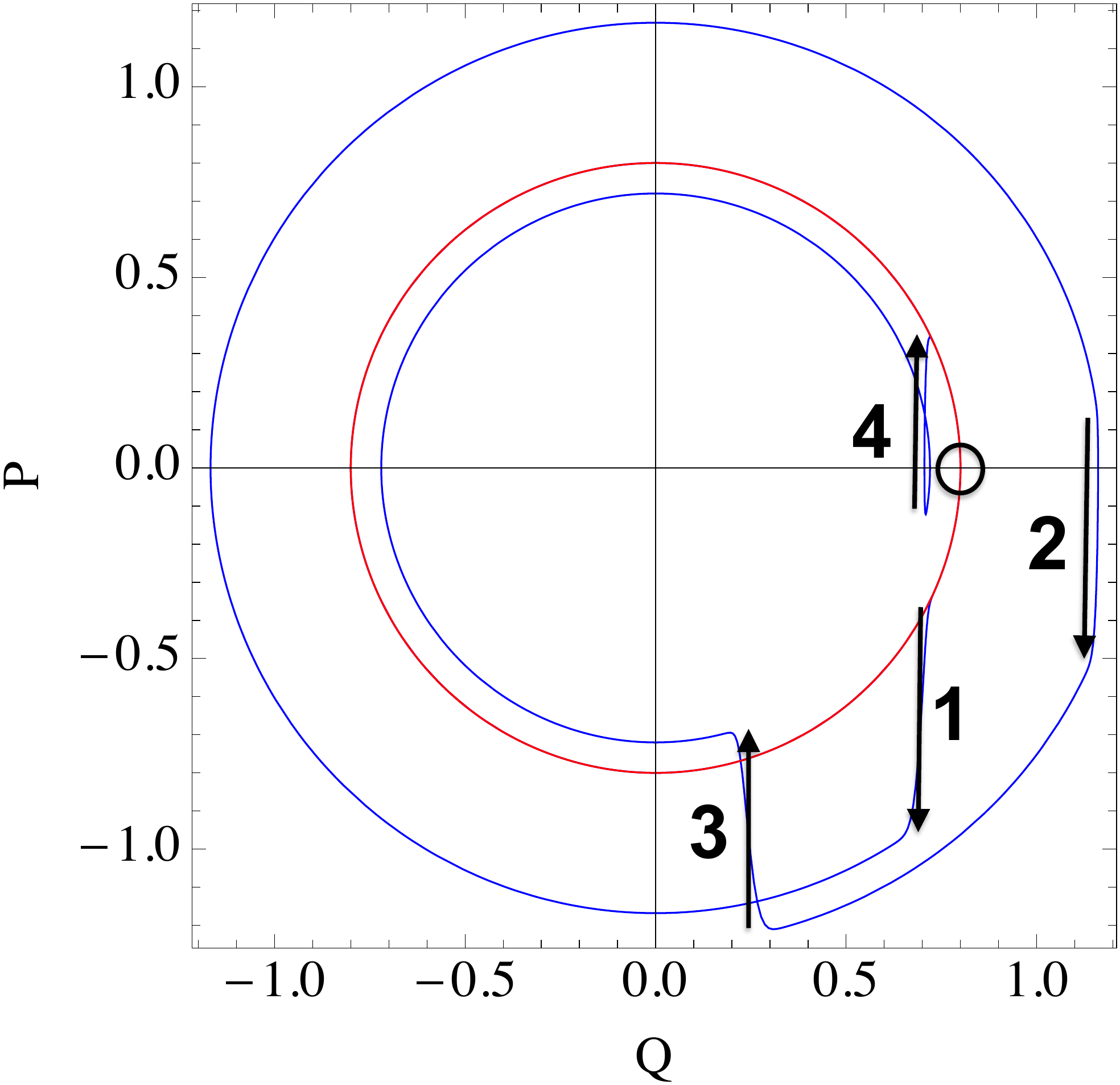}
  \includegraphics[width=0.8\linewidth]{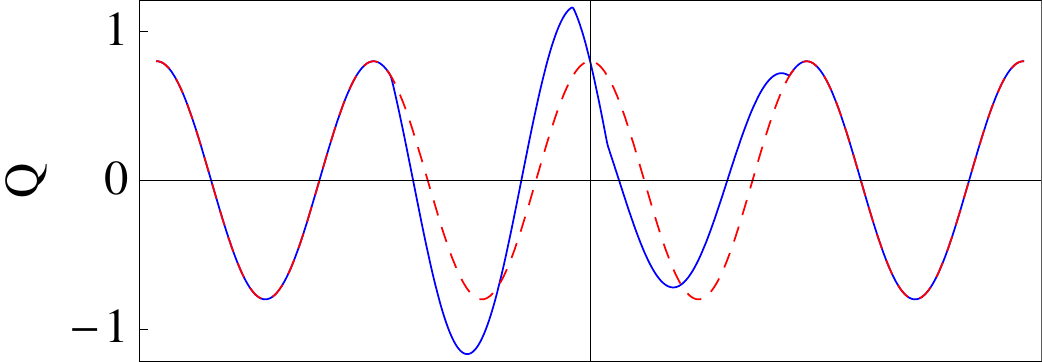}
  \includegraphics[width=0.8\linewidth]{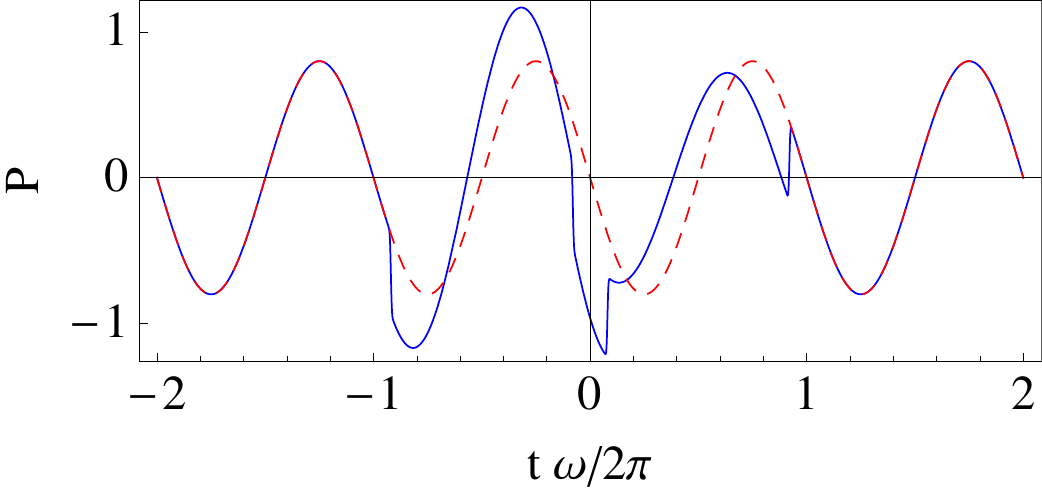}
  \caption{(color online) Effect of the four pulses~\eqref{eq:pulses} on a coherent state of a harmonic oscillator. The red dashed lines correspond to the unforced motion while the solid blue lines are subject to the kicks. (top) Phase-space plot: the trajectory starts at the point indicated by the small black circle. The arrows indicate the effect of each successive pulse. (bottom) Position, $Q$, and momentum, $P$, as a function of time, $t$. Note how both variables return to their initial orbit after the fourth kick.}
  \label{fig:pulses}
\end{figure}

We will work in the pulsed regime, with $\nu=0$ and $T\ll 2\pi/\omega,$ applying four kicks on the ions. The kicks will be spread in time according to
\begin{equation}
  \{(\Delta{k},-t_1),(\Delta{k},-t_2),(-\Delta{k},t_2),(-\Delta{k},t_1)\},
  \label{eq:pulses}
\end{equation}
with $(t_1,t_2)=(0.920, 0.080)2\pi/\omega$, see Fig.~\ref{fig:pulses}. This sequence, which takes two periods of a trap, is similar to the one developed in Ref.~\cite{garcia-ripoll03} for femtosecond resonant laser pulses, but the forcing mechanism in that reference can not be implemented with molecules because spectroscopic addressing is very hard. We emphasize that we rely instead on ultra-fast {\em off-resonant} kicks. Interestingly, as we show below, working with off-resonant kicks does not result in particularly demanding field intensities.

The momentum kick can be approximately related to the desired phase~(\ref{eq:phase-gate}) as follows
\begin{equation}
  \Delta{k} \sim T \sqrt{f_{\textrm{C}}^0 a_{\textrm{C}} f_{\textrm{T}}^0 a_{\textrm{T}}} \sim \sqrt{\phi_{\textrm{CT}}}.
\end{equation}
Requiring that the photon scattering probability remains below the small value $\varepsilon \simeq \Gamma {\Omega^2 T}/{\Delta^2} \ll 1,$ we can extract the product of the average Rabi frequency, $\Omega$, and the kick duration, $T,$ as a function of the detuning, $\Delta,$ the spontaneous emission rate of the ions, $\Gamma_{\textrm{C,T}},$ and the error tolerance, $\varepsilon.$ From here we obtain the relation
\begin{equation}
  \label{eq:3}
  \sqrt{\phi_{\textrm{CT}}} \sim \varepsilon \times \frac{\Delta}{\Gamma}\times \frac{a_{\textrm{C,T}}}{\ell},
\end{equation}
which states that we need low trap frequencies, steep light gradients and large detunings in order to acquire a large phase shift.

\begin{figure}
  \centering
  \includegraphics[width=0.9\linewidth]{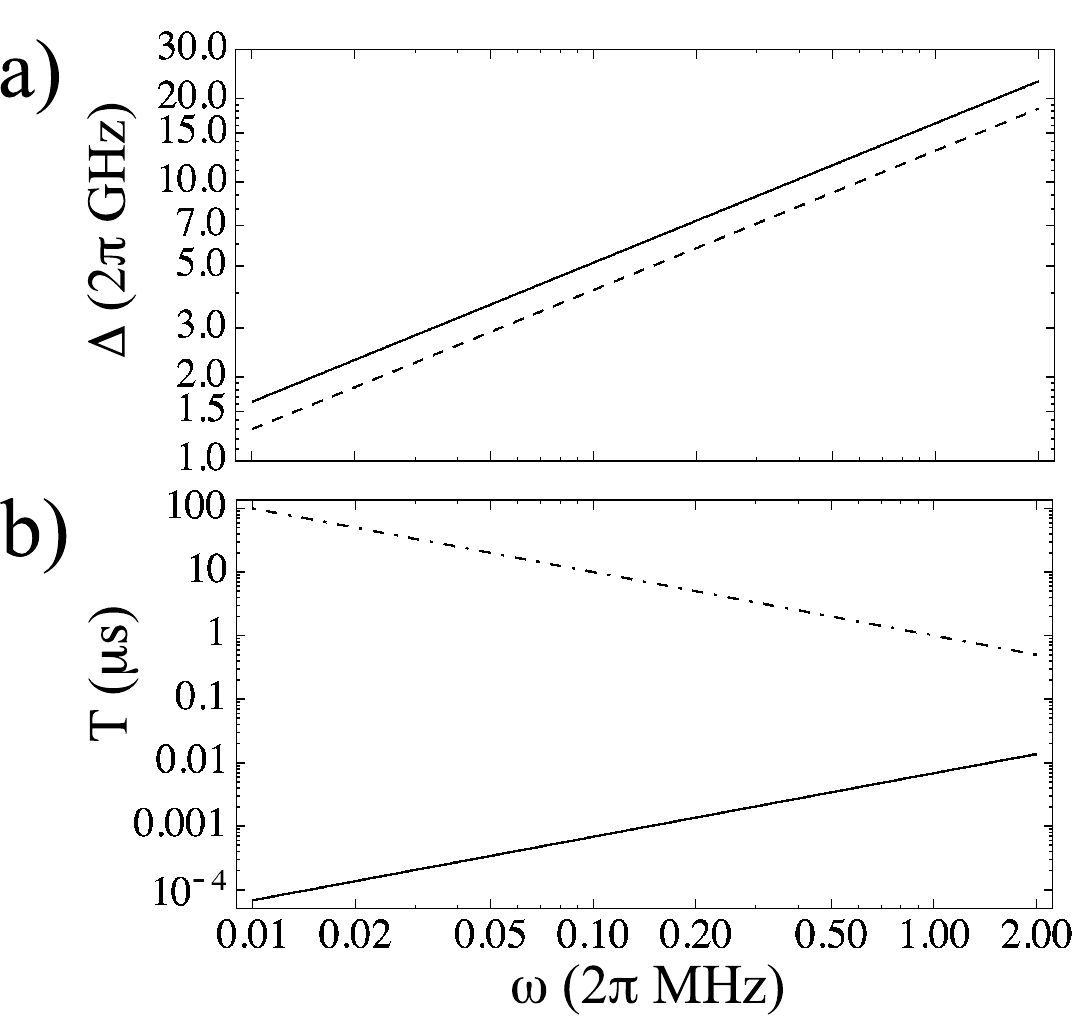}
  \caption{%
    Optical kicking protocol acting on a \Ca\ and a \Nitrogen~ion. (a) Required detunings vs. trap frequency, for a scattering error $\varepsilon=1\%$ and beam waist $\ell=5\mumeter$ (solid), and $\varepsilon=0.1\%$ in an optical lattice, $\ell=\lambda/2=397$~nm (dashed). (b) Gate pulse time (solid) and total gate duration including pauses (dash-dot), for $\varepsilon=1\%, \ell=5\mumeter$ and $1$ mW laser power. For simplicity, we assumed the same detuning for the lasers acting on the \Ca\ ($^2S_{1/2} \rightarrow {^2P}_{1/2}$, 397~nm) and
\Nitrogen\ ($\ket{X^2\Sigma_g^+, v''=0, J''=1/2, F_1} \rightarrow \ket{B^2\Sigma_u^+, v'=0, J'=3/2, F_1}$, 391~nm)
transitions.}
  \label{fig:optical}
\end{figure}

We have studied numerically the conditions to perform these quantum gates with a variety of atomic and molecular species. As an example, in Fig.~\ref{fig:optical}a we plot the required detunings for making an accurate quantum gate ($\varepsilon\sim 0.1\%,$ dashed) and for running the spectroscopy protocol ($\varepsilon=1\%,$ solid) using a setup with \Ca\ as the control ion and \Nitrogen\ as the target ion. For the atomic ion we chose the cycling transition that is excited in electron-shelving measurements, $^2S_{1/2} \rightarrow {^2P}_{1/2}$, for which existing lasers may be conveniently reused. For the molecular ion we choose the strong electronic transition $\ket{X^2\Sigma_g^+, v''=0, J''=1/2, F_1} \rightarrow \ket{B^2\Sigma_u^+, v'=0, J'=3/2, F_1}$, which would allow us to detect states that are already produced in the laboratory~\cite{tong2010}. As expected, if we need greater accuracy in the two-qubit gate without exceedingly large laser power ($\varepsilon\sim 0.1\%$), it is best to adopt an optical lattice configuration with the ion sitting on the slope, cf.~Fig.~\ref{fig:scheme}b,d. If we are only interested on spectroscopy and state determination, we may increase the error tolerance to $\varepsilon\sim 1\%$ and place the ions at half the waist from the center of a focused laser beam, cf.~Fig.~\ref{fig:scheme}a,c. In both cases the duration of the pulse is much shorter than the trap frequency, and gate times of microseconds are feasible using laser powers of milliwatts.

\subsection{Magnetic forces for Zeeman states}
\label{ssec:zeeman}

In order to discriminate Zeeman states within a molecular rovibrational level, we develop a second version of the protocol, based on the use of magnetic forces, which result from the coupling of a magnetic field gradient with the effective magnetic moment $\mu$ of the particle~\cite{mintert2001,wunderlich2011},
\begin{equation}
  f_{\textrm{T}}(t) = \frac{\partial B(t)}{\partial x} \frac{\mu}{\hbar}.
  \label{eq:force-magnetic}
\end{equation}
In this setup spontaneous emission is not a concern, but the forces are going to be typically much weaker. This may be compensated by driving the magnetic field close to resonance with one vibrational mode of the ion crystal as realized in Ref.~\cite{ospelkaus08}: for a driven force [Eq.~(\ref{eq:gaussian})] with $\nu=(1+\eta)\omega_{\textrm{com}}$ close to the center of mass mode, the phase is enhanced, $\Phi(\eta) \sim \Phi(0)/\eta,$ at the expense of a longer gate, $1/\eta$ times longer.


\begin{figure}
  \centering
  \includegraphics[width=\linewidth,clip=true]{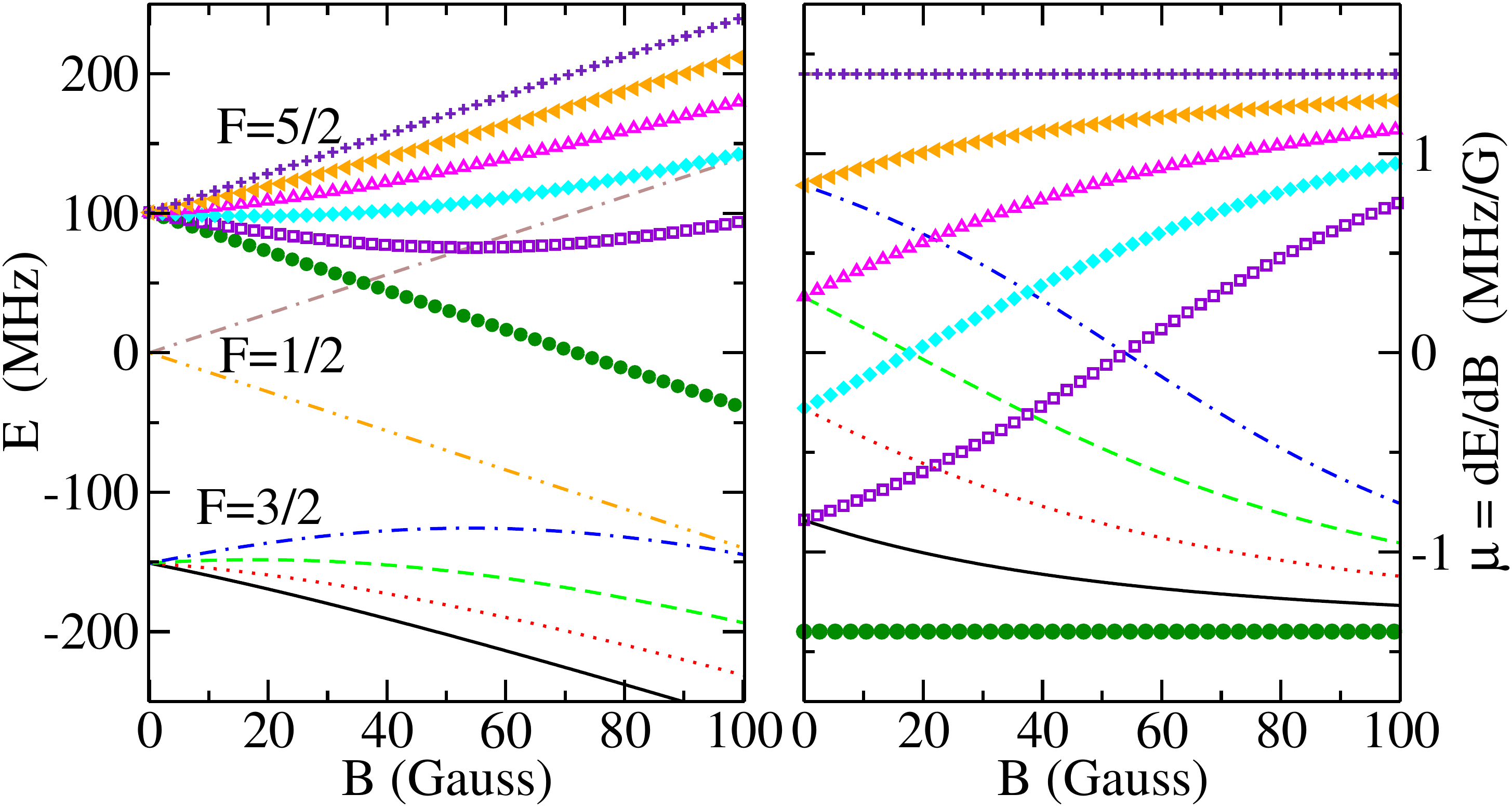}
  \caption{(color online) 
    Zeeman energy shifts of the hyperfine levels of the ground rovibronic state of \Nitrogen\ (left) and their derivatives,  $\mu = (\partial E/\partial B) /\hbar$ (right), which are needed to estimate the state dependent force    (\ref{eq:force-magnetic}).}
  \label{fig:zeeman}
\end{figure}

As a relevant, realistic example, we study the Zeeman splitting of the 
hyperfine structure of the \Nitrogen\ molecular ion in the rovibrational 
ground state of the electronic ground state potential
X~$^2\Sigma_g^+$~\cite{tong2010,tong2011}. The molecular Hamiltonian in the
presence of an external magnetic field reads~\cite{hyper-PRA,Carrington-book,Carrington-paper}
\begin{equation}
  H=H_{\textrm{rot}}+H_{\textrm{sr}}+H_{\textrm{hfs}}+H_{eqQ}+ H_{\textrm Z},
 \label{eq:molehamil}
\end{equation}
where $H_{\textrm{rot}}$ is the rotational Hamiltonian, $H_{\textrm{sr}}$ is the spin-rotation interaction, $H_{\textrm{hfs}}$ is  the magnetic hyperfine interaction, neglecting the coupling between the external magnetic field and the  nuclear spin, $H_{eqQ}$ is the nuclear electric quadrupole interaction, and finally $H_{\textrm Z}$ is the Zeeman interaction. We have calculated the lowest Zeeman levels and their respective magnetic moments, shown in Fig.~\ref{fig:zeeman} (see Appendix~\ref{sec:app-molec} for details). It is remarkable that almost all states are distinguishable by their magnetic moments, not requiring more than 10\% accuracy in its determination.

\begin{figure}[t]
  \centering
  \includegraphics[width=0.9\linewidth]{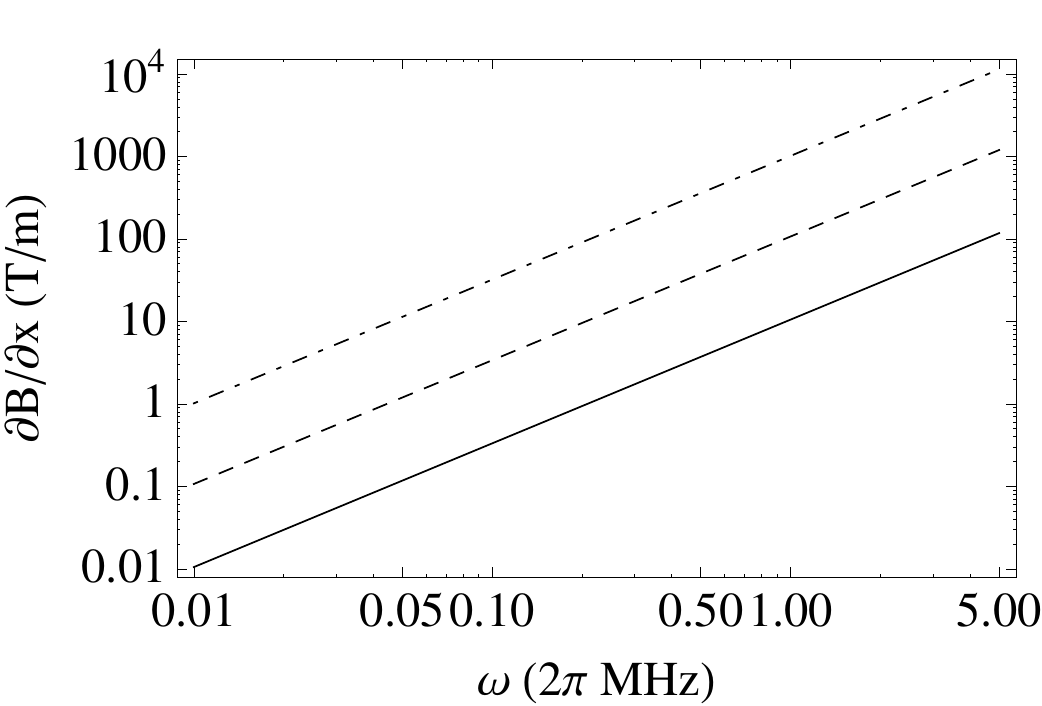}

  \includegraphics[width=0.9\linewidth]{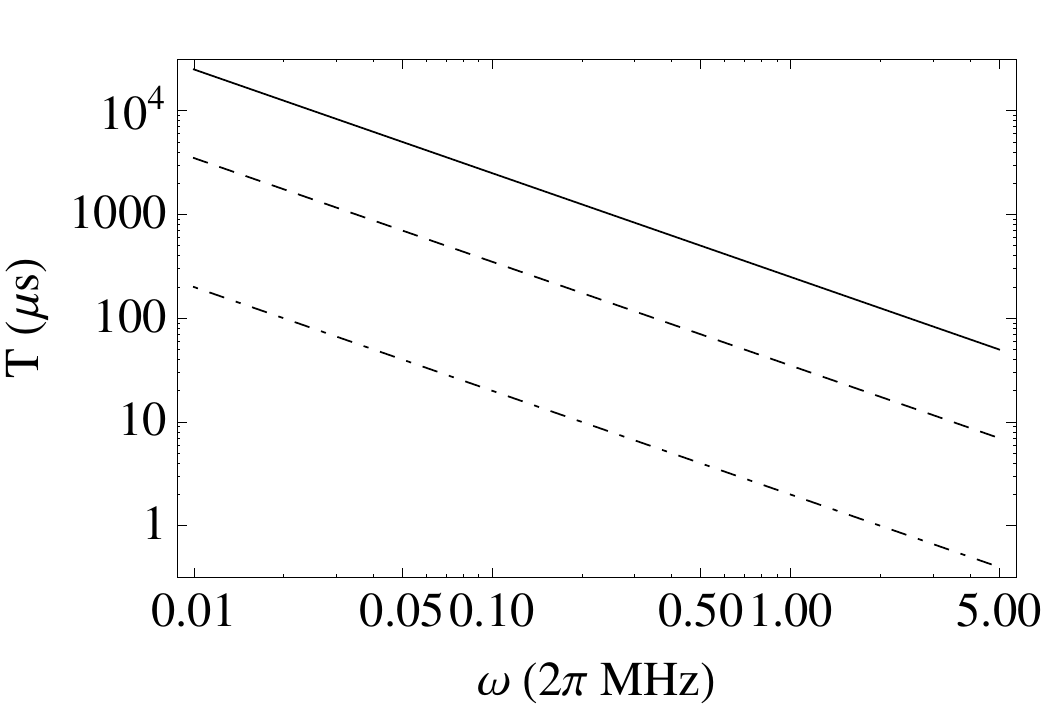}
  \caption{(top) Gradient and (bottom) time required to achieve maximum entanglement between \Ca\ and \Nitrogen\ ions, using an oscillating magnetic field gradient  $\partial B/\partial x = \exp[-(2t/T)^2] B' \cos(\nu t)$ in a harmonic trap, $\omega,$ for drivings $\nu/\omega=0,\,1.1$ and $1.01$ (dash-dotted, dashed, solid).}
  \label{fig:mag-fields}
\end{figure}

Using the magnetic moments one can easily compute the field intensities required to perform interferometry using a \Ca\ and a \Nitrogen\ ion, as a function of the trap frequency. As shown in Fig.~\ref{fig:mag-fields}, a trap with $\omega = 2\pi\times574$~kHz requires a magnetic field gradient of 10 T/m for a 250$\mu$s gate using a driving $\nu=1.01\omega,$ values which do not seem disparate~\cite{welzel11,ospelkaus08}. However, note that for the same duration 
there exists another configuration for $\nu=0$ which produces the same gate at a lower value of $B'=\partial B/\partial x,$ using a smaller trap frequency. More precisely, if we can make a gate with field gradient $B'$ using $\nu=(1+\eta)\omega_{\textrm{com}}$ in a time $T_{\nu}\sim 5\pi/\eta\omega$, we can perform the same gate using a smaller trap frequency $\omega_{\nu=0} \propto \eta^{2/3}\omega_\nu$ and a significantly shorter time 
$T_{\nu=0} \propto \eta^{1/3}T_\nu$. 
This setup may be advantageous for larger traps, where achieving a large $B'$ may be difficult and would otherwise require long gate times; using in this case $\nu=0$ would reduce the gate time, thus avoiding heating problems~\cite{turchette00,safavi2011}.

\subsection{Opto-magnetic protocol}
\label{ssec:opto-mag}

Both of the preceding methods may hit an important problem with certain molecules which due to their mass or their internal structure respond more weakly to the external optical and magnetic forces. For such cases, we propose a third protocol that combines the application of ultrashort laser pulses on the control ion with arbitrary optical or magnetic forces on the target. Driving the control ion with a sequence of $N$ femtosecond $\pi$-pulses~\cite{garcia-ripoll03} we will be able to inject an almost arbitrary amount of momentum $\Delta k_{\textrm{C}} \sim N\times 2\pi/\lambda$ to the system. The enhanced pushing capability of the pulse train compensates the smaller kicks, $\Delta k_{\textrm{T}},$ suffered by the target particle, as the total phase is proportional to the product, $\phi_{\textrm{CT}} \sim \Delta k_{\textrm{C}}\Delta k_{\textrm{T}}.$ This simple idea may be used in an ultrafast combined opto-magnetic protocol to lower the magnetic field requirements by various orders of magnitude

\section{Conclusions and outlook}
\label{sec:conclusions}
 
In conclusion, we have introduced a spectroscopy protocol based on quantum logic that is applicable to a wide range of ions without accessible cycling transitions, such as molecular ions.
We have presented numerical results, based on original molecular structure calculations, showing that the requirements to implement this protocol with molecular ions of current interest are within experimental reach.
This protocol, which combines magnetic and optical forces, enables the realization of measurements and entanglement on hybrid atom-molecule ion systems without the need for cooling to the trap vibrational ground state, a significant improvement over previous methods. This constitutes a new approach to high-resolution spectroscopy of cold molecular ions, and opens the door towards novel hybrid quantum computation schemes.

\acknowledgments

This work is supported by Spanish MICINN Projects FIS2009-10061, FIS2010-22064-C02-02 and CTQ2007-62898-BQU, CAM research consortium QUITEMAD S2009-ESP-1594, the Swiss National Science Foundation through the National Centre of Competence in Research ``Quantum Science and Technology'', ESF COST Action IOTA (MP1001), a FP7 Marie Curie fellowship (IEF-2009-251913 MOLOPTLAT), and a JAE CSIC Fellowship.

\appendix

\section{Calculation of the total phase acquired}
\label{sec:app-phase}
We sketch here the derivation of the phase acquired by the two-ion state due to the forces applied on the ions, Eq.~\eqref{eq:phaseCT}. We start by recalling the total phase acquired by a state of a harmonic oscillator of frequency $\omega$ driven by a force $f(t)$ for a time $T$~\cite{garcia-ripoll05}:
\begin{equation*}
  \phi = \frac{1}{2\hbar^2} \textrm{Im} \int_0^T dt \int_0^t d\tau e^{i\omega(t-\tau)} f(t) f(\tau) .
\end{equation*}
For a time dependence as in Eq.~\eqref{eq:gaussian}, and assuming a sufficiently long pulse so that the motional state is restored to its initial value, $T \geq 5\pi/\textrm{min}{|\nu-\omega|}$, this phase takes the form
\begin{equation}
  \phi[\omega,f] = \sqrt\frac{\pi}{2} \left( \frac{f^0 a}{\hbar} \right)^2 \frac{\omega T}{8(\omega^2-\nu^2)} .
  \label{eq:phase1}
\end{equation}
with $a=\sqrt{\hbar/(m\omega)}$ the harmonic oscillator length.

For the case of a two-ion system, we have to take into account that a force applied on one of the ions will affect the two common modes, center-of-mass (com) and stretch (str). Applying forces $F_1$ and $F_2$ on the two ions, the forces on the com and str modes are
\begin{align*}
  F_{\textrm{com}} &= F_1 + F_2 , \\
  F_{\textrm{str}} &= \frac{1}{m_1 + m_2}\left( m_2 F_1 - m_1 F_2 \right) .
\end{align*}
The phase acquired is then given by Eq.~\eqref{eq:totalphase}. Using the result~\eqref{eq:phase1}, we obtain that the only contribution to $\Phi$ that depends on $F_1 F_2$ is
\begin{align*}
  \Phi &= F_1 F_2 \sqrt{\frac{\pi}{2}} \frac{T}{8(m_1+m_2)\hbar}
  \left(  \frac{2}{\omega_{\textrm{com}}^2 -\nu^2}
    - \frac{2}{\omega_{\textrm{str}}^2 -\nu^2}
  \right) ,
\end{align*}
which is readily rewritten as Eq.~\eqref{eq:phaseCT} with $f^0_{\textrm{C,T}} = F_{1,2}/\hbar$.

\section{Calculation of \Nitrogen\ hyperfine levels}
\label{sec:app-molec}

\begin{table*}[bth]
  \caption{Values of the molecular Hamiltonian parameters used in the calculations. $B_e$ and $D_e$ are taken from Ref.~\cite{wu2007} and are here given in cm$^{-1}$, while the other parameters are taken from~\cite{hyper-PRA} and given in MHz.}
  \begin{tabular}{l|cccccccc}
    Parameter & $B_e$ & $D_e$ & $\gamma$ & $\gamma_N$ & $b_{\textrm F}$ & $c_{\textrm{dip}}$ & $c_I$ & $eqQ$ \\ \hline
    Value & 1.9223897 & $5.9758 \times 10^{-6}$ & 276.92253 & $-3.9790 \times 10^{-4}$ & 100.6040 & 28.1946 & 0.01132 & 0.7079 \\
  \end{tabular}
  \label{tab:params}
\end{table*}

The complete molecular Hamiltonian for a $^2 \Sigma$ molecule such as N$_2^+$, including the hyperfine structure and the corresponding splittings under a magnetic field $B$, involves several angular momenta: the rotational angular momentum $\bfm{N}$, the (total) electronic spin $\bfm{S}$, and the total nuclear spin $\bfm{I}$. These momenta are coupled within a Hund's case (b) scheme, i.e., $\bfm{J}=\bfm{N}+\bfm{S}$ and  $\bfm{F}=\bfm{J}+\bfm{I}$. The terms in Eq.~\eqref{eq:molehamil} are given by~\cite{hyper-PRA,Carrington-book,Carrington-paper}
\begin{itemize}
\item The rotation of the molecule including the centrifugal distorsion,
  \begin{equation}
    H_{\textrm{rot}}=B_{e}\bfm{N}^{2} - D_{e}\bfm{N}^{4} .
  \end{equation}

\item The fine structure spin-rotation interaction, 
  \begin{equation}
    H_{\textrm{sr}}= \gamma_{\textrm{sr}} \bfm{N} \cdot \bfm{S} .
  \end{equation}

\item The magnetic hyperfine Hamiltonian that couples the nuclear spin with the electronic spin and molecular rotation, 
  \begin{equation}
    H_{\textrm{hfs}}=H_{I S} + H_{I N}  .
  \end{equation}
  The first contribution can be written as a sum of the Fermi contact interaction, $H_{\textrm F} = b_{\textrm F}  \bfm{I} \cdot \bfm{S}$, and the dipolar term, $H_{\textrm{dip}}=-\sqrt{10}T^{(1)}(\bfm{I})\cdot T^{(1)}(\bfm{S},C^2)$, in spherical tensor notation (see Ref.~\cite{Carrington-paper} for details). The second part, $H_{I N}$, is similar to the spin-rotation term but considering the nuclear spin instead, $ H_{I N}= c_I \bfm{I} \cdot \bfm{N}$.

\item The nuclear electric quadrupole interaction, which can be expressed using spherical tensor notation as
  \begin{equation}
    H_{eqQ}=e\sum_{i=1,2}T^{(2)}(Q_i)T^{(2)}(\nabla\cdot\bfm{E}_i) ,
  \end{equation}
  with $e$ the electron charge.

\item Finally, for the Zeeman interaction we include only the coupling of the electronic spin with the magnetic field
  \begin{equation} 
    H_{\textrm Z }=g \mu_B B S_z
  \end{equation}
  where $g$ is the electron gyromagnetic factor (taken $g=2$), $B$ is the magnetic field, $\mu_B$ is the Bohr magneton, and $S_z$ is the component of the electronic spin along the quantization axis that is taken in the direction of the external field. Any other magnetic contributions will be neglected as they are of much less importance.
\end{itemize}

We work within the Hund's case (b) and select the basis $\ket{ NSJIFM_{F} }$ to represent the Hamiltonian. The nuclear spin of $^{14}$N equals $I_1=1$, and the calculations have been carried out for the symmetry block which involves even values of both the rotational ($N=0,2,4,\ldots$) and the total nuclear spin ($I=0,2)$. 
For zero magnetic field, our results are in complete agreement with those appearing in~\cite{hyper-PRA}.
We have also checked that the calculations reported here are converged just including $N=0,2$ in the Hamiltonian matrix.

With the values for the parameters of the molecular Hamiltonian reported in Table~\ref{tab:params}, the matrix elements read explicitly as follows~\cite{hyper-PRA}:
\begin{widetext} 
\begin{itemize}
\item The rotational term:
  \begin{eqnarray}
    \label{ec29a}
    \bra{ N'S'J'I'F'M'_{F} } H_{\textrm{rot}} \ket{ NSJIFM_{F} }=  
    \delta_{NN'} \delta_{SS'} \delta_{JJ'} \delta_{II'} \delta_{FF'} 
    \delta_{M_{F}M'_{F}} \left( B_{e}N(N+1) -D_{e}[N(N+1)]^2 \right) .
  \end{eqnarray}

\item The fine structure, spin-rotation term:
  \begin{eqnarray}
    \label{ec29b}
    \bra{ N'S'J'I'F'M'_{F} } H_{\textrm{sr}} \ket{ NSJIFM_{F} } = 
    \delta_{NN'} \delta_{SS'} \delta_{JJ'} \delta_{II'} \delta_{FF'} 
    \delta_{M_{F}M'_{F}}  \frac{ \gamma_{\textrm{sr}}}{2} 
    \left[J(J+1)-N(N+1)-S(S+1)\right] .
  \end{eqnarray}
  Here, the coupling constant is defined by $\gamma_{\textrm{sr}}=\gamma + \gamma_{N} N(N+1)$.

\item The magnetic hyperfine structure terms:
\begin{align}
  \langle N'S' & J'I'F'M'_{F} | H_{\textrm F} \ket{ NSJIFM_{F} }
  = 
  \delta_{NN'} \delta_{SS'} \delta_{II'} \delta_{FF'} \delta_{M_{F}M'_{F}} 
  b_{\textrm F}(-1)^{F+I+J+J'+N+S+1} 
  \nonumber \\
  &
  \times \sqrt{I(I+1)(2I+1)} \sqrt{S(S+1)(2S+1)}  \sqrt{(2J'+1)(2J+1)} 
  \sixj{I}{J}{J'}{I}{F}{1} \sixj{S}{J}{J'}{S}{N}{1}  ;
  \label{ec30} 
  \\
  \langle N'S' & J'I'F'M'_{F} | H_{\textrm{dip}} \ket{ NSJIFM_{F} }
  = 
  \delta_{SS'} \delta_{II'} \delta_{FF'} \delta_{M_{F}M'_{F}} 
  c_{\textrm{dip}} (-1)^{F+I+J+N'+1} 
  \nonumber \\
  &
  \times \sqrt{30 \, I(I+1)(2I+1)S(S+1)(2S+1)}  
  \sqrt{(2J'+1)(2J+1)(2N+1)(2N'+1)}
  \nonumber \\
  &
  \times \sixj{I}{J}{J'}{I}{F}{1} \ninej{N'}{N}{2}{S}{S}{1}{J'}{J}{1}
  \threejm{N'}{0}{2}{0}{N}{0}  ;
  \label{ec31} 
  \\
  \langle N'S' & J'I'F'M'_{F} | H_{IN} \ket{ NSJIFM_{F} }
  = 
  \delta_{NN'} \delta_{SS'} \delta_{II'} \delta_{FF'} \delta_{M_{F}M'_{F}} 
  c_{I}(-1)^{F+I+J'+J+N+S+1}
  \nonumber \\
  &
  \times \sqrt{I(I+1)(2I+1)} \sqrt{N(N+1)(2N+1)}  
  \sqrt{(2J'+1)(2J+1)} \sixj{J}{I}{1}{F}{J'}{I} \sixj{N}{J'}{1}{S}{N}{J}  ,
  \label{ec34}
\end{align}
with $\threejm{\cdot}{\cdot}{\cdot}{\cdot}{\cdot}{\cdot}$,$\sixj{\cdot}{\cdot}{\cdot}{\cdot}{\cdot}{\cdot}$, $\ninej{\cdot}{\cdot}{\cdot}{\cdot}{\cdot}{\cdot}{\cdot}{\cdot}{\cdot}$, 3-$j$, 6-$j$, and 9-$j$ symbols, respectively. All matrix elements couple functions with  $\Delta J=0,\pm1$. In addition, the dipolar interaction, $H_{\textrm{dip}}$, mixes states with $\Delta N=0,\pm2$. The values for $b_{\textrm F}$,
$c_{\textrm{dip}}=t + t_{N} N (N+1)$, and $c_{I}$ are given in Table~\ref{tab:params}. We have neglected the centrifugal distortion term $t_{N}$ because of its very small value.

\item The nuclear quadrupole term, that allows for a number of off-diagonal elements ($\Delta J= 0, \pm1,\pm2$; $\Delta I=0,\pm2$; $\Delta N = 0,\pm2$):
  \begin{align}
    \bra{ N'S'J'I'F'M'_{F} } & H_{eqQ} \ket{ NSJIFM_{F} }= 
    \delta_{SS'} \delta_{FF'} \delta_{M_{F}M'_{F}} 
    \frac{eqQ}{2} \frac{(-1)^{I}+(-1)^{I'}}{2}(-1)^{F+I'+2J}  (-1)^{2I_{1}+S+2N'}
    \nonumber \\
    &
    \times 
    \sqrt{(2I+1)(2I'+1)(2J+1)(2J'+1)(2N+1)(2N'+1)}
    \nonumber \\
    &
    \times \sixj{I'}{J}{2}{F}{I}{J'} 
    \sixj{I_{1}}{I}{2}{I_{1}}{I_{1}}{I'} \sixj{N'}{J}{2}{S}{N}{J'}
    \threejm{N'}{0}{2}{0}{N}{0} 
    \threejm{I_{1}}{-I_{1}}{2}{0}{I_{1}}{I_{1}}^{-1}
    \label{ec32} 
  \end{align}
  where $I_1$ is the nuclear spin of one of the nuclei composing the (homonuclear) molecule, and the  value of $eqQ$ is also from Ref.~\cite{hyper-PRA}.

\item Finally the Zeeman term~\cite{Carrington-paper}:
  \begin{align}
    \bra{ N',S', J',I',F',M'_{F} } & H_{\textrm Z} \ket{ N,S,J,I,F,M_{F} }=
    \delta_{NN'}  \delta_{SS'} \delta_{II'}  g \mu_B  B
     (-1)^{F'-M'_{F}}  (-1)^{2J'+N+S+F+I}
    \nonumber \\
    &
    \times   
    \sqrt{(2F'+1)(2F+1)} 
    \sqrt{(2J'+1)(2J+1)} 
    \sqrt{S(S+1)(2S+1)}
    \nonumber \\
    &
    \times
    \threejm{F'}{-M'_{F}}{1}{0}{F}{M_{F}}
    \sixj{F'}{J}{J'}{F}{I}{1} \sixj{J'}{S}{S}{J}{N}{1} \:.
    \label{we-15} 
  \end{align}
\end{itemize}
\end{widetext}

\bibliography{bibliomol}

\end{document}